\documentclass[twocolumn,preprintnumbers,amsmath,amssymb,nofootinbib
,superscriptaddress]{revtex4}
\usepackage{color}
\usepackage{hyperref}
\usepackage{graphicx}
\usepackage{color}
\usepackage{graphicx,graphics}

\newcommand{\be}{\begin{equation}}  
\newcommand{\ee}{\end{equation}}  
\newcommand{\bear}{\begin{eqnarray}}  
\newcommand{\eear}{\end{eqnarray}}  
\newcommand{\ba}{\begin{array}}  
\newcommand{\ea}{\end{array}}

\newskip\humongous \humongous=0pt plus 1000pt minus 1000pt

\newif\ifdtup

\def\oldreffmt#1{\rlap{[#1]} \hbox to 2\parindent{}}

\def\figfmt#1{\rlap{Figure {#1}} \hbox to 1in{}}  
  
\def\ie{\hbox{\it i.e.}{}}	  
\def\eg{\hbox{\it e.g.}{}}

\def\beq{\begin{equation}}  
\def\eeq{\end{equation}}  
\def\bea{\begin{eqnarray}}  
\def\eea{\end{eqnarray}}

\def\bq{\begin{quote}}  
\def\eq{\end{quote}}

\relax  

\newdimen\tdim  
\tdim=\unitlength  
\def\bar{\overline}

\begin{document}

\preprint{FERMILAB-PUB-16-035-T}

\title{Scale-Independent Inflation and Hierarchy Generation }

\author{Pedro G. Ferreira}
\email{pedro.ferreira@physics.ox.ac.uk}
\affiliation{Astrophysics, Department of Physics\\
University of Oxford, Keble Road\\
Oxford OX1 3RH\\$ $}
\author{Christopher T. Hill}
\email{hill@fnal.gov}
\affiliation{Fermi National Accelerator Laboratory\\
P.O. Box 500, Batavia, Illinois 60510, USA\\$ $}
\author{Graham G. Ross}
\email{g.ross1@physics.ox.ac.uk}
\affiliation{Rudolf Peierls Centre for Theoretical Physics, \\
University of Oxford, 1 Keble Road\\
Oxford OX1 3NP\\$ $}

\date{\today}

\begin{abstract}
We discuss models involving two scalar fields coupled to classical gravity that satisfy the general
criteria: (i) the theory has no mass input parameters, (ii) 
classical scale symmetry is broken only through 
$-\frac{1}{12}\varsigma \phi^2 R$ couplings where $\varsigma$ departs from the special conformal value of $1$;
(iii) the Planck mass is dynamically generated by the vacuum expectations values (VEVs) of the scalars (iv) there is a stage of viable inflation associated with slow roll in the two--scalar potential; (v) the final vacuum has a small to vanishing
cosmological constant and an hierarchically small ratio of the VEVs and the ratio of the scalar masses to the Planck scale.   This assumes the paradigm of classical scale symmetry
as a custodial symmetry of large hierarchies.
\end{abstract}

\maketitle


The discovery of the weakly interacting Brout-Englert-Higgs (BEH) boson, coupled with the absence of 
significant evidence for physics beyond the Standard Model, has stimulated a re-evaluation of the possible
 explanations of the hierarchy problem. In the Standard Model (SM) of the strong and electroweak interactions, 
which has no fundamental input mass scale
other than the BEH mass, an apparent hierarchy problem  arises that is due to the  additive 
quadratically  divergent radiative corrections to the mass squared of the BEH boson.
However, in the pure Standard Model
the quadratic divergences are an artifact of the introduction of a mass scale cut-off
in momentum space \cite{bardeen}.  In the context of field theory, the coefficients of relevant operators have to be 
renormalised and the theory is defined ultimately by observable renormalised coefficients. In this case neither 
the quadratically divergent radiative correction to the BEH mass nor the mass counter-term is measurable and only 
the renormalised mass is physically meaningful.  If one maintains scale invariance broken
only explicitly by the various trace anomalies and spontaneously to generate
 the BEH boson mass, then the latter must be viewed as multiplicatively renormalized 
since no quadratic divergence arises in the trace anomaly.  
This has further led to the proposal of classically-scale-invariant models that contain the SM, in 
which the electroweak scale is generated through spontaneous breaking of scale invariance via
Coleman-Weinberg mechanism \cite{CW,general}. 
 
It has been suggested that scale invariance might even apply at the quantum level 
through ``endogenous" renormalisation 
which requires that the regulator mass
scale, $\mu$,  associated with quantum loops in dimensional regularization,   
is itself generated by a moduli field\footnote{{For a recent discussion see \cite{Ghil} and, in the context of the model discussed below see \cite{Higgsinf1}.}}. 
Alternatively, one can always introduce an arbitrary cut-off scale $\Lambda$, \eg,  
by way of momentum space cut-off or Pauli-Villars regularization, but then renormalize the 
theory at a renormalization scale given by a moduli field to remove
the $\Lambda$ dependence\footnote{It is easy to see that if one subtracts at some mass scale $M$ 
that is specified externally to the 
defining field theory action, then the trace anomaly arises as the variation of the renormalized 
action wrt $\ln(M)$.  In replacing
the subtraction scale $M$ by an actual field $\chi$ that is part of the defining
action of the theory, there is no residual trace anomaly; the trace anomaly is simply absorbed into the improved 
stress tensor itself, which then remains traceless.}. 
However we will not explore this possibility here, concentrating on whether it is possible 
to build a viable scale invariant theory broken only spontaneously and via the trace anomaly. 

Of course a complete theory must include gravity and, if one is to maintain classical scale invariance, 
it is necessary to do so in a way that generates the Planck scale through spontaneous breaking of the 
scale invariance such as occurs in the Brans Dicke theory of gravity \cite{brans}. The inclusion of gravity 
means there are additional additive divergent contributions to the BEH mass but these, too, are unphysical 
and should be absorbed in the renormalised mass which, {as before, is multiplicatively renormalised due to the underlying scale invariance and thus avoids the hierarchy problem.}

{A problem with the scale independent approach occurs if there are massive states coupled to the BEH scalar for then there are large finite calculable corrections to the Higgs mass. In the Standard Model the presence of the Landau pole associated with the $U(1)$ gauge group factor indicates  that the SM becomes strongly interacting at the scale associated with the Landau pole.   It is common to assume that there will be massive bound states associated with this strong interaction that will couple significantly to the BEH boson and create the ``real" hierarchy problem. One possible way to evade this is to embed the SM  in a theory with no Abelian gauge group factor that does not have a Landau pole \cite{Giudice:2014tma}. This must be done close to the electroweak scale to avoid introducing the hierarchy problem via new massive states and leads to a profusion of new states that may be visible at the LHC. However the Landau pole in the SM lies above the Planck scale where gravitational effects cannot be neglected and it is far from clear clear what the physics above the Landau pole will be and whether it indeed reintroduces the hierarchy problem. For the same reason we did not insist on the absence of a Landau pole in the model considered here.}
 
{Similarly it is possible  that, when gravity becomes strong, it leads to massive states that generate the real hierarchy problem. Of course there are black holes that can carry SM gauge group charges and couple to the BEH boson. In general  such states do not give rise to an hierarchy problem due to their form factor suppression. It is possible that microscopic black holes exist that do not have such form factor suppression but this is not firmly established and, as with the Landau pole problem, we chose to ignore this possibility here.}

In this paper we construct a spontaneously broken scale-free model that includes gravity. As such, there is no physical meaning 
to the vacuum expectation value (vev) of a single scalar field and only dimensionless ratios are measurable. 
A mimimal model  capable of generating an hierarchy  requires the introduction of two scalar fields, 
$\phi$ and $\chi$ coupled to gravity in the form: 
\begin{eqnarray}
S&=&-\int d^4 x\sqrt{-g}[\frac{1}{12}\alpha \phi^2 R+\frac{1}{2}\nabla_\mu\phi\nabla^\mu \phi \nonumber\\
&& \qquad + \frac{1}{12}\beta \chi^2 R+\frac{1}{2}\nabla_\mu\chi\nabla^\mu \chi + W(\phi, \chi)]
\label{S}
\end{eqnarray}
where: $
W(\phi,\chi)   =  \lambda \phi^4 + \xi\chi^4+\delta\phi^2\chi^2 
$.
This theory has no input mass scales, is conformally invariant if $\alpha=\beta=1$ and  is
invariant under independent $\phi\rightarrow \pm\phi,\;\;\chi\rightarrow \pm\chi$.  
\begin{figure}[htbp]
\vskip-0.1in
\begin{flushleft}
\includegraphics[width=8.9cm]{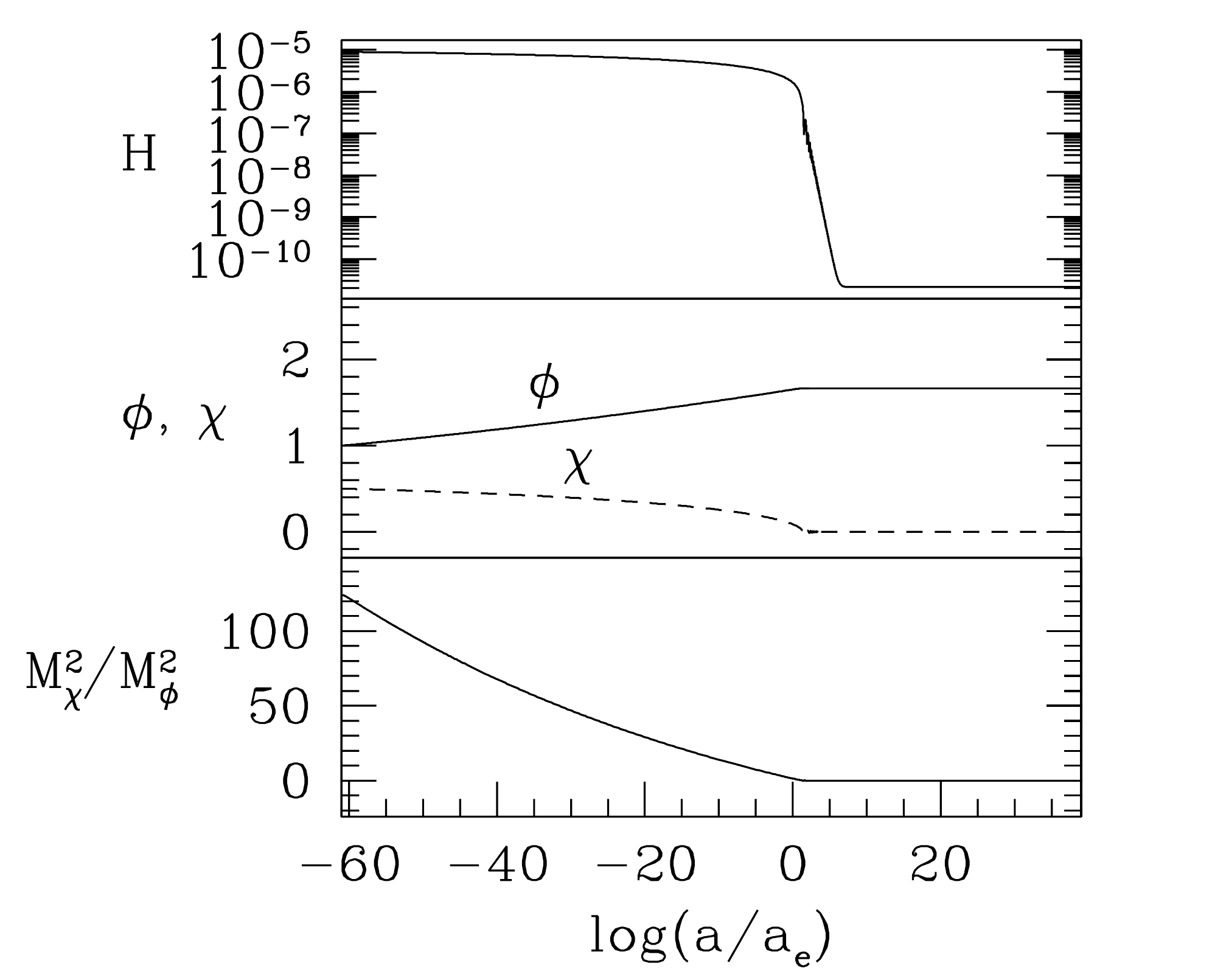} 
\end{flushleft}
\vspace{-20pt}
\caption{{Plot of the Hubble parameter, $H$, $\phi$, $\chi$ and the ratio of the two components of the effective Planck mass, $M^2_\phi$ and $M^2_\chi$, as a function of $a$; we have normalized the x-axis to the scale factor at the end of inflation, $a_e$. }}
\label{figure_back}
\end{figure} 

{The theory has remarkable properties that we illustrate for one representative choice of parameters 
($\alpha$, $\beta$, $\lambda$, $\xi$, $\delta$) in Figure \ref{figure_back}. At early times it has a 
period of inflation during which, as we will show later on, observationally viable spectra of scalar 
and tensor perturbations can be generated. Furthermore, it has an infra-red (IR) fixed point set by 
ratios of the coupling constants and which is radiatively stable to quantum corrections and during which 
the universe can undergo accelerated expansion.}

{In the context of unimodular gravity\footnote{The unimodular constraint does not play a role during the inflationary stage.} references \cite{Higgsinf, Higgsinf1} provide  seminal studies of the model. These studies concentrate on the $\xi=O(1)$ case  in which the field $\chi$ may be interpreted as the Higgs, in turn requiring $\beta=O(10^5)$ to produce  ``Higgs inflation".  }

{In this paper we extend the analysis to cover other values of the parameters. By way of motivation we note that in the context of the hierarchy problem it is important that there should be no heavy states significantly coupled to the Higgs. In this case it has been argued \cite{Allison:2014hna} that the solution to the strong CP problem requires the introduction of the axion and, in the context of this model, the most economical solution is to identify the axion with a component of the $\chi$ field. However then the coupling $\xi$ must be small to avoid the introduction of a low-lying Landau pole.
A second difference is that we determine the inflationary solution in the ``Jordan" frame of eq(\ref{S}) whereas the analysis of reference \cite{Higgsinf, Higgsinf1} was performed in the Einstein frame. Our analysis has the advantage that it has a simple analytic solution in the slow-roll region, clarifying the origin of the structure of the model. Finally, the IR fixed point structure of the model studied here differs from that in \cite{Higgsinf, Higgsinf1} where the unimodular constraint introduces an explicit cosmological constant.}

In the Jordan frame the field equations immediately follow from eq.(\ref{S}):
\be
M^2 G_{\alpha\beta}=T^\phi_{\alpha\beta}+T^\chi_{\alpha\beta} 
- g_{\alpha\beta}W(\phi,\chi)
\label{sign}
\ee
where
\begin{eqnarray}
T^\phi_{\alpha\beta}&=&\left(1-\frac{\alpha}{3}\right)\nabla_\alpha\phi\nabla_\beta\phi
+\left(\frac{\alpha}{3}-\frac{1}{2}\right)g_{\alpha\beta}\nabla_\mu\phi\nabla^\mu\phi\nonumber\\
&&-\frac{\alpha}{3}\phi\nabla_\alpha\nabla_\beta\phi+\frac{\alpha}{3} g_{\alpha\beta}\phi\Box\phi \nonumber \\
T^\chi_{\alpha\beta}&=&\left(1-\frac{\beta}{3}\right)\nabla_\alpha\chi\nabla_\beta\chi
+\left(\frac{\beta}{3}-\frac{1}{2}\right)g_{\alpha\beta}\nabla_\mu\chi\nabla^\mu\chi\nonumber\\
&&-\frac{\beta}{3}\chi\nabla_\alpha\nabla_\beta\chi+\frac{\beta}{3} g_{\alpha\beta}\chi\Box\chi 
\end{eqnarray}
and:
\begin{eqnarray}
 \Box\phi-\frac{\alpha}{6}\phi R-{\partial W\over\partial\phi}=0, \ \ \ 
 \Box\chi-\frac{\beta}{6}\chi R-{\partial W\over\partial\chi}=0.
\end{eqnarray}

{The effective planck mass, $M^2=M^2_\phi+M^2_\chi$ 
(where $M^2_\phi=-\alpha\phi^2/6$ and $M^2_\chi=-\beta\chi^2/6$) is time varying during during the 
inflationary period (when $M^2_\phi\ll M^2_\chi$) but constant during the late time accelerated 
expansion phase (when $M^2_\phi\gg M^2_\chi$), obeying current constraints on gravitational physics. 
To obtain the normal form of the Einstein equations at late times, $M^2$ must be positive and 
therefore at least one of the coefficients $\alpha$  or $\beta$ must be negative, inconsistent with 
the conformally invariant choice. However the resultant theory is still scale-independent  \cite{related}. }

Taking the 
trace of the Einstein field equations we have:
\begin{eqnarray}
-M^2R&=&(\alpha-1)\nabla_\mu\phi\nabla^\mu\phi+(\beta-1)\nabla_\mu\chi\nabla^\mu\chi
\nonumber\\ & &
+\alpha\phi\Box\phi+\beta\chi\Box\chi-4W
\end{eqnarray}
which determines the Ricci scalar.

We now restrict the analysis to study the cosmological evolution for a Friedmann Robertson Walker 
(FRW) metric, $g_{\alpha\beta}=(-1, a^2\delta_{ij})$. 
The FRW equation is given by:
\be
H^2-\frac{D}{3M^2}H-\frac{\rho_T}{3M^2}=0 
\label{H2}
\ee
where $H\equiv{\dot{a}}/{a}$ is the Hubble parameter, 
$D=\alpha\phi\dot{\phi}+\beta\chi\dot{\chi}$ and $\rho_T=\dot{\phi}^2/2+\dot{\chi}^2/2+W$.
The evolution equations for $\phi$ and $\chi$ can be uncoupled to give:
\begin{eqnarray}
\left( \begin{array}{c}
\Box\phi\\
\Box\chi\end{array} \right)=\frac{1}{\rm K}\left( \begin{array}{cc}
1+\frac{\beta^2\chi^2}{6M^2} &-\frac{\alpha\beta\phi\chi}{6M^2}\\
-\frac{\alpha\beta\phi\chi}{6M^2} &1+\frac{\alpha^2\phi^2}{6M^2}  \end{array} \right)\left( \begin{array}{c}
{\cal S}_\phi\\
{\cal S}_\chi\end{array} \right) \label{fulleqs}
\end{eqnarray}
where ${\rm K} = 1+(\alpha^2\phi^2+\beta^2\chi^2)/(6M^2)$ and:
\begin{eqnarray}
{\cal S}_\phi&=&\alpha(\alpha-1)\frac{\phi\dot{\phi}^2}{6M^2}+\alpha(\beta-1)\frac{\phi\dot{\chi}^2}{6M^2}
\nonumber\\&+&\frac{4\alpha\phi}{6M^2}W+{\partial W\over \partial \phi} \nonumber \\
{\cal S}_\chi&=&\beta(\beta-1)\frac{\chi\dot{\chi}^2}{6M^2}+\beta(\alpha-1)\frac{\chi\dot{\phi}^2}{6M^2}\nonumber\\
&+&\frac{4\beta\chi}{6M^2}W+{\partial W\over \partial \chi}
\label{boxeq}
\end{eqnarray}

As advertised, this theory has an infrared fixed point which can be found
by setting $\ddot{\phi}=\dot{\phi}=\ddot{\chi}=\dot{\chi}=0$ leading to: 
\begin{eqnarray}\label{paireqs}
\bar{\cal S}_\phi&\equiv &-4\alpha\frac{\phi}{\alpha\phi^2+\beta\chi^2}W
+{\partial W\over \partial\phi}=0\nonumber \\
\bar{\cal S}_\chi&\equiv &-4\beta\frac{\chi}{\alpha\phi^2+\beta\chi^2}W+{\partial W\over \partial \chi}=0 \label{eigen}
\end{eqnarray}
Note that $ \phi\bar{\cal S}_\phi + \chi\bar{\cal S}_\chi=0$ is automatically satisfied
since our full potential, $W(\phi,\chi)$, 
 is classically scale invariant:
$
{\delta{W}}/{\delta \ln\phi}
+ {\delta W}/{\delta \ln\chi}=4W$
This guarantees that  nontrivial solutions generally exist in
the ratio of the VEV's of $\phi$ and $\chi$ given by:
\begin{eqnarray}
\frac{\langle\chi_0\rangle^2}{\langle\phi_0\rangle^2}=\frac{4\lambda\beta-2\alpha\delta}{4\alpha\xi-2\beta\delta}
\end{eqnarray}
One can readily show that this is an IR stable fixed point so that $\langle\phi_0\rangle,\;\langle\chi_0\rangle$ are the IR vevs of the scalar fields. Note that it is only dimensionless ratios of VEVs that are physical. The absolute value of a VEV, not determined by the static equations, is not measurable.

We are interested in the case that $\langle\phi_0^2\rangle\gg \langle\chi_0^2\rangle$ so that, at late times, a large hierarchy develops. To have an hierarchically light ``matter" sector also requires that the $\chi$ mass should be small relative to the Planck scale and this in turn requires that the $\chi$ mass contribution coming from the  $\delta \phi^2\chi^2$ term should be hierarchically small relative to the Planck mass, {\it i.e.} $\delta\le \langle\chi_0^2\rangle/\langle\phi_0^2\rangle$. Finally if the cosmological constant at late times is small then this requires a fine-tuning of the parameters in $W$ such that it is (or is close to)  a perfect square. Furthermore, we need $\lambda\le \langle\chi_0^4\rangle/\langle\phi_0^4\rangle$ which, in the absence of a ${\alpha\over 12}\phi^2 R$ term,  is natural because $\phi$ is shift symmetric in the limit the small parameters vanish. Thus the radiative corrections to the small parameters can only be gravitational in origin (we will discuss these corrections later in this letter).

What happens to the scale factor in the IR?  For static scalar fields the FRW equation, Eq. \ref{H2}, implies:
\be
3M^2\left({\dot{a}\over a}\right)^2=W=(\lambda+\xi \mu^4+\delta\mu^2)\phi_0^4
\label{1.2a}
\ee
(where 
 $\mu^2\equiv {\langle\chi_0\rangle^2}/{\langle\phi_0\rangle^2}$) and we can define an effective cosmological 
constant $\Lambda_{\rm eff}=(\lambda+\xi \mu^4+\delta\mu^2)\phi_0^2/(\alpha+\beta\mu^2).$  
 {With the ordering of the couplings discussed above $\Lambda_{eff}\le \xi\chi_0^4/M^2$. To obtain zero cosmological constant requires fine tuning of the couplings corresponding to the potential having the form of a perfect square.}

{This theory is equivalent to a multi-scalar 
Jordan-Brans-Dicke theory of gravity with a potential \cite{brans,Damour1992,Berti2015}. Current constraints on Brans-Dicke theories from Shapiro time delay measurements are particularly stringent and a naive application to this theory leads to $\alpha <2.5\times 10^{-5}$. However, the scale invariance of the theory implies that  a change in the Planck mass will be compensated by a corresponding change in massive objects that cancel the effect so that the bound does not apply.} 


A remarkable feature of the scale-independent structure, that we see in Fig \ref{figure_back}, is that it readily 
leads to an inflationary era. Non-minimally coupled models of inflation have been looked at before \cite{Acetta,Fakir,Komatsu,Shafi}. 
Multifield, non-minimal models have also been looked at in some detail, with a particular focus on models with an explicit 
Planck mass \cite{Kaisermultiple} or perfectly (or almost perfect) conformal invariance (with $\alpha=\beta=1$) \cite{Linde}. 

{However this case is characteristically different, with no explicit Planck mass and the slow-roll condition resulting from a cancellation of terms due to the scale invariance of non-gravitational sector.}
To understand its inflationary regime, it useful to rewrite Eq. \ref{fulleqs} in terms of $M^2_\phi$ and $M^2_\chi$.  In the regime where  $W\simeq \xi \chi^4$, Eqs \ref{fulleqs} gives us:
\begin{eqnarray}
\frac{d}{dN}\left(\begin{array}{c}M^2_\phi
\\ M^2_\chi
\end{array} \right)=\frac{4}{3}\frac{M^2_\phi(M^2_\phi+M^2_\chi)}{(\alpha-1)M^2_\phi+
(\beta-1)M^2_\chi}\left( \begin{array}{c}
(1-\beta)\alpha\\
(\alpha-1)\beta\end{array} \right)
\label{slowroll}
\end{eqnarray}
where $N=\ln a$. {Slow-roll results in the $\beta\gg\alpha$ regime where $M^2_\chi\gg M^2_\phi$ because the scale invariant 
form of the scalar potential results in a cancellation of the large $\partial W\over \partial \chi$ term in eq(\ref{boxeq}) 
so that the rhs of eq(\ref{slowroll}) is proportional to $M_{\phi}^2$.  Solving this equation gives the inflationary solution 
$M^2_\phi=M^2_{E}e^{\nu N}$ and
$M^2_\chi=M^2_{ E}\left[1+\gamma\left(1-e^{\nu N}\right)\right]$} where $\nu=-4\alpha/3$ and $\gamma=\beta(1-\alpha)/\alpha(1-\beta)$, 
and we have  $N=0$ at the end of inflation when $M^2_{\phi }=M^2_{\chi }=M^2_E$. We have checked that this solution is a 
superb approximation to the numerical solution to Eqs \ref{fulleqs}.

With our analytical solution in hand,  assuming that at the beginning of inflation we have $\phi\sim\chi\sim \Phi_I$, 
we find that the total number of e-folding during inflation is {$N_{\rm tot}=-(1/\nu)\ln [(1+\gamma)/(\beta/\alpha+\gamma)]$.} 
This allows us to determine the value of the effective Planck mass today as a function of $M_I=-\frac{1}{6}\alpha\Phi_I^2$ through $M^2_{E}\simeq M_I^2e^{\nu N_{\rm tot}}$.
 If $\alpha, \beta \ll 1$ we have that $M^2_{E}\simeq M_I^2$ while being possible to have $N_{\rm tot}\rightarrow \infty$.

{We can also calculate the predictions for the inflationary observables \cite{Baumann}. The standard procedure, in the case of single field models is to calculate the slow parameters in the Einstein frame; following \cite{Higgsinf} we will do so here although effects arising from the multi dynamics may change our results somewhat. In the Einstein frame (which we denote with  a tilde over all quantities, e.g. $\tilde X$)  we have that the Hubble rate is given by
${\tilde H}^2({\tilde N})\simeq (36\xi/\beta^2)M^4_\chi/(3M^2_\chi-M^2_\phi)$ which we use to determine the slow roll parameters,
${\tilde \epsilon}=-{\tilde H}'$ and ${\tilde \eta}={\tilde \epsilon}-{\tilde \epsilon}'/2{\tilde \epsilon}$, and then calculate the tensor to scalar ratio, $r=16{\tilde \epsilon}$ 
and the scalar spectra index, $n_s=1+2{\tilde \eta}-4{\tilde \epsilon}$. We then find the expressions:
\begin{eqnarray}
&r=\frac{64\alpha^2}{9\zeta}\frac{e^{\nu N_e}}{(e^{\nu N_e}-1)^2}\label{r} \\
&n_s-1=-\nu\frac{e^{\nu N_e}+1}{e^{\nu N_e}-1} \label{n_s}
\end{eqnarray}
where $\zeta=\beta/(\beta-1)$ and $N_e$ is the number of e-foldings before inflation.  In order to obtain fluctuations of the observed magnitude we need $\xi/\beta^2=O(10^{-10})$. For Higgs inflation $\xi=O(1)$ so one must have $\beta=O(10^5)$. Here we explore smaller values of $\beta$ which will require correspondingly smaller values of $\xi$.
\begin{figure}[htbp]
\vskip-0.1in
\begin{flushleft}
\includegraphics[width=8cm]{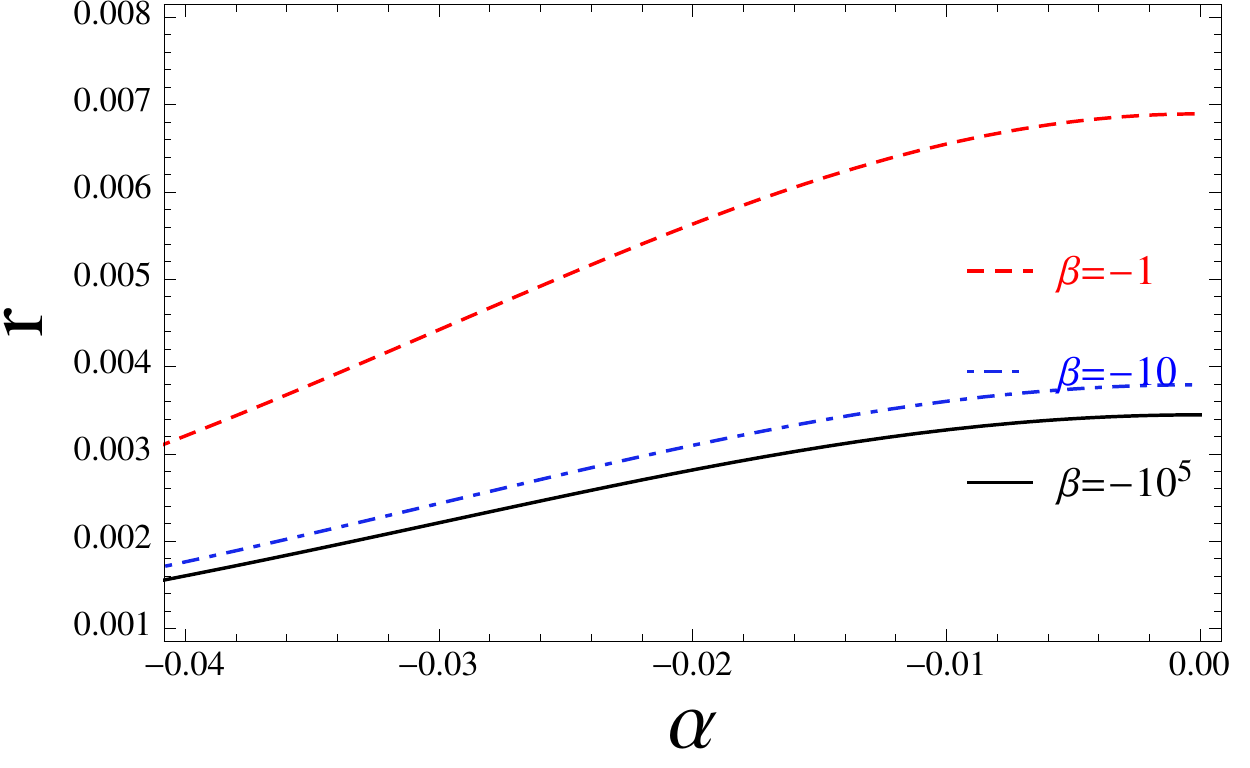} 
\includegraphics[width=8cm]{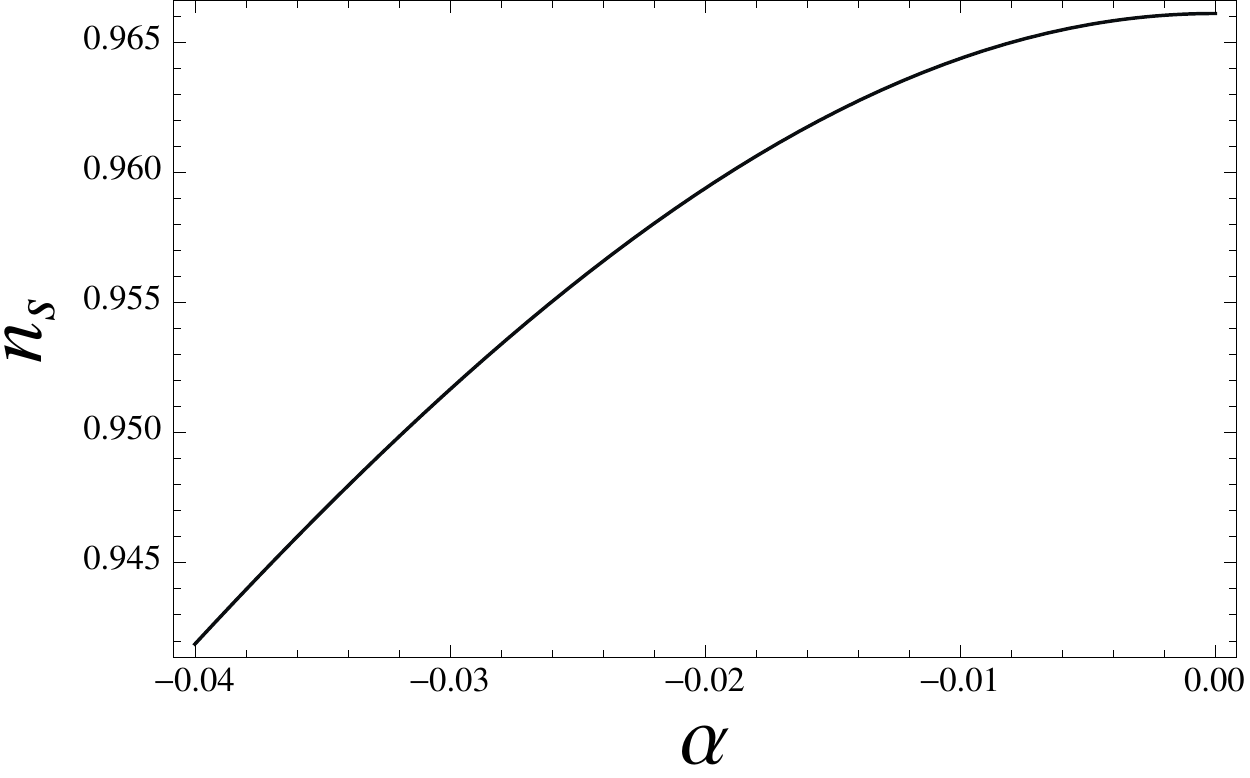} 
\end{flushleft}
\vspace{-20pt}
\caption{{Plot of the tensor to scalar ratio, $r$, and the the scalar spectra index, $n_s$, for a range of $\alpha$ and $\beta$. }}
\label{fluct}
\end{figure}}

{Figure \ref{fluct} plots $r$ and $n_s$ for a range \footnote{In the large $\beta$ limit our expressions reproduce almost exactly the numerical values obtained in \cite{Higgsinf}.} of $\beta$. It is straightforward to obtain ($r$, $n_s$) consistent with the Planck measurements 
\cite{Planck2015}, i.e. $r\le 0.1$ and $n_s\sim 0.96$; indeed, typical values of $r$ range from $10^{-3}$ to $10^{-2}$. Future B-mode constraints will further tighten bounds on $r$, leading to bounds on $\alpha$ and $\beta$. }

{This analysis has assumed that only a single field is active. Being a two field system, $\Phi$, there are possible additional  isocurvature  fluctuations and non-negligable non-Gaussian effects \cite{Gordon} that are proportional to $\eta_{\perp}.\delta\Phi_\perp$ where $\eta_\perp$ and  $\delta \Phi_\perp$ are the components of the slow roll vector,  $\overrightarrow\eta$, and the field perturbations orthogonal to the background field trajectory respectively. In the slow roll regime one may see from eq(\ref{slowroll}) that the ratio $M_\phi^2/M_\chi^2$ is field independent implying $\eta_\perp$ vanishes, being an attractor of the scale invariant theory\cite{Higgsinf1}, thus justifying the assumption. }

{The generation of an hierarchy requires that the choice of parameters in the tree level Lagrangian is also hierarchical and it is important to check whether this choice is stable against radiative corrections. The choice $\lambda\ll \delta\ll\xi$ is stable against non-gravitational corrections because in the limit that $\lambda$ and $\delta$ vanish there is an enhanced shift symmetry $\phi\rightarrow\phi+c$. This implies that non-gravitational corrections to $\delta$ are proportional to $\delta$ while the corrections to $\lambda$ are proportional to $\delta^2$ or $\lambda$, both being perturbatively small. 
The gravitational corrections have been studied in detail in reference \cite{Higgsinf1} and we do not repeat the discussion here. Calculating the radiative calculations using dimensional regularisation as an example of endogenous renomalisation it was shown that the model results discussed here are essentially unchanged by gravitational corrections. }

{While the model is very simple, it provides a basis to extend the Standard Model to include gravity in a scale invariant theory. Reference \cite{Higgsinf} identified the $\chi$ field with the Higgs scalar and so that the inflationary era is Higgs inflation. However this is not the only possibility. As we commented above it may be advantageous to identify $\chi$ as the field giving rise to the axion solution to the strong CP problem.}  Of course the SM states should have hierarchically small coupling to the $\phi$ field but such small couplings will again be radiatively stable due to the enhanced symmetry when the couplings are zero.

We have shown that a simple
two-scalar model coupled to gravity can satisfies the general
criteria: (i) the theory has no mass input parameters, \ie, is classically scale
invariant. {One can readily see that this model possesses a conserved current of
the form  $j_\mu =(1-\alpha)\phi\partial_\mu \phi+(1-\beta)\chi\partial_\mu \chi$.
This current arises upon combining eqs.(4,5) to eliminate $R$ 
and it is covariantly conserved, $D^\mu j_\mu = 0$ and it plays an important role in the dynamics which will be explored in subsequent work ref.(22); }(ii) scale symmetry is broken only through 
the scalar coupling to the Ricci scalar which depart from the special conformal value of $-1/6$;
(iii) the Planck mass is dynamically generated by the scalar VEV's (iv) there is a viable stage of inflation
associated with slow roll in the two--scalar potential; (v) the final vacuum has a small to vanishing
cosmological constant and an hierarchical ratio between the Planck scale and the scalar mass scale.   Our analysis assumes the paradigm of scale symmetry as a custodial symmetry of large hierarchies. We will present generalizations of this scheme to multi-scalar theories as well as the inclusion of SM states
and expand the formal implications elsewhere \cite{FHR2}.



\vskip .2 in
\noindent
 {\bf Acknowledgements}

\vspace{0.1in}
We thank W. A. Bardeen, C. Burrage, J. Dunkley, A.Lukas, J. Rubio, S. Sarkar, M. Shaposhnikov, D. Sloan, L. Stein, K. Yagi  for discussions.
Part of this work was done at Fermilab, operated by Fermi Research Alliance, 
LLC under Contract No. DE-AC02-07CH11359 with the United States Department of Energy. PGF acknowledges support from STFC, the Beecroft Trust and the Higgs Centre in Edinburgh.
 

\end{document}